# Associations between park features, park satisfaction and park use in a multi-ethnic deprived urban area

Hannah Roberts, Ian Kellar, Mark Conner, Christopher Gidlow, Brian Kelly, Mark Nieuwenhuijsen, Rosemary McEachan

**Abstract**

Parks are increasingly understood to be key community resources for public health, particularly for ethnic minority and low socioeconomic groups. At the same time, research suggests parks are underutilised by these groups. In order to design effective interventions to promote health, the determinants of park use for these groups must be understood.

This study examines the associations between park features, park satisfaction and park use in a deprived and ethnically diverse sample in Bradford, UK. 652 women from the Born in Bradford cohort completed a survey on park satisfaction and park use. Using a standardised direct observation tool, 44 parks in the area were audited for present park features. Features assessed were: access, recreational facilities, amenities, natural features, significant natural features, non-natural features, incivilities and usability. Size and proximity to the park were also calculated.

Multilevel linear regressions were performed to understand associations between park features and (1) park satisfaction and (2) park use. Interactions between park features, ethnicity and socioeconomic status were explored, and park satisfaction was tested as a mediator in the relationship between park features and park use.

More amenities and greater usability were associated with increased park satisfaction, while more incivilities were negatively related to park satisfaction. Incivilities, access and proximity were also negatively associated with park use. Ethnicity and socioeconomic status had no moderating role, and there was no evidence for park satisfaction as a mediator between park features and park use.

Results suggest diverse park features are associated with park satisfaction and park use, but this did not vary by ethnicity or socioeconomic status. The reduction of incivilities should be prioritised where the aim is to encourage park satisfaction and park use.

**Highlights**

- Associations between park features, use and satisfaction in a deprived, ethnically diverse sample are examined
- More amenities, greater usability and fewer incivilities are associated with higher park satisfaction
- Increased access and more incivilities are associated with reduced use
- Ethnicity and socioeconomic status do not moderate relationships between park features, satisfaction and use
- Park satisfaction does not mediate the relationship between park features and use

# 1 Introduction

Green space is increasingly understood to be a valuable environmental resource in health promotion (Nieuwenhuijsen, Khreis, Triguero-Mas, Gascon, & Dadvand, 2017; WHO, 2016). Recent systematic reviews have highlighted the association between exposure to green space and improvement in both mental (Bowler, Buyung-Ali, Knight, & Pullin, 2010; Gascon et al., 2015) and physical health outcomes (Twohig-Bennett & Jones, 2018). It is suggested that green space can encourage physical activity, improve air quality, reduce stress and encourage social interaction (Hartig, Mitchell, De Vries, & Frumkin, 2014). Furthermore, several studies have demonstrated that the relationship between green space and health may be strongest in ethnic minority and low socioeconomic status groups (Maas, Verheig, Groenewegen, de Vries, & Spreeuwenberg, 2006; Mitchell & Popham, 2007, 2008). However, recent research shows that green spaces are under-utilised, particularly by these groups (Cohen, Han, Derose, et al., 2016; Evenson, Wen, Hillier, & Cohen, 2013). This poses an important public health challenge whereby the use of green space should be encouraged for those who could benefit the most, yet currently are among those who use it the least.

Key determinants of green space use are thought to be structural factors such as size and proximity (Bedimo-Rung, Mowen, & Cohen, 2005). For example, in a nationwide study of 174 parks across 25 cities in the US, Cohen et al. (2016) found that every additional acre of park land was associated with a 9% increase in park use. In addition, Coombes et al. (2010) conducted a survey of nearly 7000 adults in England and found a significant decline in frequency of park use as objective proximity increased. It has been suggested that the unequal spatial distribution of green space is a contributing factor in the reduced level of use among ethnic minority and low socioeconomic status groups (Floyd, Taylor, & Whitt-Glover, 2009; Jennings, Johnson Gaither, & Gragg, 2012). Several studies have shown that these groups have reduced access to and provision of green space (Ferguson, Roberts, McEachan, & Dallimer, 2018; Wolch, Byrne, & Newell, 2014). However, other studies have shown access and provision is at least comparable or even improved (Barbosa et al., 2007; Kessel et al., 2009; Timperio, Ball, Salmon, Roberts, & Crawford, 2007).

Park quality may also play a key role in determining park use: for instance, Kaczynski et al. (2008) observed 33 parks in Canada, and found that while size and proximity were not significant predictors of park use, the number of features was significant. Furthermore, specific features that encourage park use or park-based physical activity have been identified, including playgrounds, paved trails, basketball courts, water features, shelter and picnic areas (Baran et al., 2014; Costigan, Veitch, Crawford, Carver, & Timperio, 2017; Kaczynski et al., 2009; Rung, Mowen, Broyles, & Gustat, 2011; Shores & West, 2008). On the other hand, some features have been shown to discourage use. These include incivilities, such as litter and vandalism, and poor quality of playing surfaces (Gobster, 2002; McCormack, Rock, Toohey, & Hignell, 2010; van Hecke et al., 2018).

Some studies have shown the relationship between park features and use varies by ethnicity and socioeconomic status (Hughey et al., 2016; Kaczynski et al., 2014; Vaughan, Colabianchi, Hunter, Beckman, & Dubowitz, 2018). For example, Kaczynski et al. (2014) demonstrated that fitness stations and skate parks were related to park use only for those on a low income, while playgrounds, baseball fields and basketball courts were associated with park use only for Black users. In addition, many studies have reported that ethnic minority and low socioeconomic groups have reduced quality parks available to them in comparison to White and more affluent groups (Bruton & Floyd, 2014; Rigolon, Browning, & Jennings, 2018; Suminski et al., 2012; Vaughan et al., 2013). In England, it has been demonstrated that ethnic minority groups and those living in deprived areas more often perceive their local green space as poor quality, and this is linked to reduced usage patterns compared to the White British population (CABE, 2010). Altogether, these results highlight the need to design or modify parks that are in line with the needs of the community and avoid a 'one size fits all' approach. In doing so, the benefits of green space can be realised to their full potential across diverse communities.

Current research concerning park features and use is concentrated in North America. However, it is difficult to make comparisons between North America and the UK in this regard. First, the spatial context is different: US cities are more sprawling and more dependent on cars for travel (Richardson et al., 2012). Second, the ethnic and racial context is also different: North American studies tend to focus on Hispanic and African American populations (Engelberg et al., 2016; Vaughan et al., 2018), whereas the major ethnic minority groups in the UK are of South Asian origin (Indian, Pakistani, Bangladeshi) (Roe, Aspinall, & Thompson, 2016).

In this study we address this research gap by examining a sample of mothers within a birth cohort located in an ethnically diverse and deprived city in the north of England, of which 50.1% of participants are South Asian (Wright et al., 2013). Also, limited research in the US has shown that satisfaction with the quality of neighbourhood public space is interrelated with use of green and social spaces (Hadavi & Kaplan, 2016). Therefore, in this study we build on this research and suggest park satisfaction as a mediator of the relationship between park quality and park use. As a result, we conduct two multilevel models to understand the associations between park features, park satisfaction and park use. Interactions are explored by ethnicity and socioeconomic status. We also test park satisfaction as a potential mediator.

Overall, the primary aim of this study was to explore the associations between park features, park satisfaction and park use. Secondary aims were to explore whether ethnicity and socioeconomic status moderate the relationship between park features, park use and satisfaction, and to explore whether park satisfaction mediated the relationship between park features and park use.

**2 Methods**

Study design and setting

This study utilised a multi-method design. We collected bespoke data from the Born in Bradford (BiB) cohort, a longitudinal cohort of 12,453 mothers. Participants in this cohort were recruited during pregnancy between 2007 to 2011. The aim of the cohort is to examine how genetic, nutritional, environmental, and social factors impact on health and development of children. A full description of the cohort has been reported elsewhere (Wright et al., 2013).

In addition, an observational audit of 44 parks within Bradford to record park features was conducted by a team of researchers from 15th June to 3rd July 2015 using the Natural Environment Scoring Tool (NEST) (Gidlow et al., 2018). Figure 1 shows a map of the audited parks.

Bradford is situated in West Yorkshire in the north of England. With a population of just over half a million, the population density is 14.3 persons per hectare (Office for National Statistics (ONS), 2017). It has high levels of deprivation with 32.6% of neighbourhoods in the district in the most deprived 10% of neighbourhoods nationally on the English Index of Multiple Deprivation 2015 (Department for Communities and Local Government, 2015). Twenty-five percent of the population of Bradford are South Asian (Indian, Pakistani, Bangladeshi). Just over 80% of this group are Pakistani (ONS, 2017). Within the whole BiB cohort, 50.1% of participants are of South Asian origin (Wright et al., 2013).

Participants

Potential participants (sample size (n) =843) were those included in a sub-study of the BiB cohort who responded to a survey that included questions on their child's park use and satisfaction, between June 2013 and June 2015. The sample was limited to those who had complete data and those who had listed an audited park as their most visited park during the summer months, resulting in a total of n=652 included in the analysis (see Figure 2 for a flow diagram).

Variables

*Park-level variables*

*Park quality*

Respondents were asked "which park(s) does your child play in most frequently during summer?"; this was repeated for winter. It was possible to name up to two parks for each question. In total, 224 unique parks across all four possible answers were identified by respondents. Responses were collated by seasons and selected for auditing based on those most frequently identified by respondents. All parks that were nominated in both the summer and winter months and were reported more than once in at least one season were audited, ensuring that data was available for parks that were used consistently throughout the year.

In total, 44 parks were selected. The majority of selected parks were local neighbourhood parks with a variety of features such as a children's play area, walking paths, benches and open green space for sports or relaxing. Some were smaller with fewer features and limited green space; some were much larger with large open green spaces, more developed sports facilities, and a variety of other facilities such as toilets and cafes.

Parks were audited in-situ using the Natural Environment Scoring Tool (NEST) (Gidlow et al., 2018). The tool lists 47 items categorised into eight domains: access, recreational facilities, amenities, natural features, non-natural features, incivilities, significant natural features, and usability. The items in each domain are shown in Table 1.

Most items are assessed on both presence and quality simultaneously, so that a higher score indicates the item is present and of good quality; zero means the feature is not present in the park. Some items ask for presence only, for example 'good view point', with presence (=1) or absence (=0) as responses. The usability domain scores the space on how suitable the park appears to be for various activities, e.g. walking, socialising, children's games ("not suitable", "somewhat suitable", "suitable", "very suitable").

Two independent observers assessed each park using the NEST. The level of agreement between observers was calculated, ICC = 0.90. Item scores were entered by observers into Microsoft Excel and compared. Any disagreements between observers were resolved by selecting the higher of the two scores provided, i.e. presence was the default. Items were recoded to indicate presence (=1) or absence (=0) of each feature. Usability was dichotomised (does not support the activity= 0 and supports activity = 1). Scores were summed to produce an overall score for each domain.

*Park size*

All audited parks were mapped in ArcGIS (ESRI, 2018) and park size was calculated in hectares.

*Individual-level variables*

*Park use*

Respondents were asked on how many days and for how long on average during the weeks and the weekend their child used the park(s) they listed as frequently using, for both summer and winter. Outlying or implausible values were removed. An average annual index of use was calculated for each participant by multiplying the number of days by the number of minutes for the week and the weekend and summing for each season, then averaging between the seasons. Park use was measured in average minutes per week over the course of the year.

*Park satisfaction*

Park satisfaction was assessed by asking participants to rate their satisfaction with the park(s) they listed on a Likert type scale (1=very dissatisfied to 5=very satisfied). This was found to be not normally distributed, and so was collapsed to a 3-point scale (whereby 1-3 were aggregated), with higher ratings indicating greater satisfaction.

*Park proximity*

Straight-line distance between the centroid of the respondent' postcodes and the boundary of their most visited park in summer was calculated.

*Socio-demographics*

    *Ethnicity*

Ethnicity was self-reported in the BiB baseline questionnaire. Responses were categorised into three groups: White British, Pakistani and a diverse mixed 'Other' due to the large proportion of White British and Pakistani respondents (combined total of 85%). The final category represents a mix of ethnicities including White Other (3.1%), Mixed Race (1.8%), Black (2.1%), Indian (3.7%), Bangladeshi (2.4%) and 'Other' (1.9%).

    *Socioeconomic status*

Socioeconomic status (SES) was measured at an individual and area level. Individual-level indicators were education, measured by highest educational qualification (0 = maximum of 5 GCSEs, 1 = A level equivalent or above), and financial status ('How well would you say you or you and your husband/partner are managing financially these days?') (0= struggling financially, 1 = not struggling financially). The Index of Multiple Deprivation (IMD) (2010) was used as an indicator of area-level deprivation (Department of Communities and Local Government, 2011). The indicator is available at the lower super output area (LSOA) level, the lowest administrative geographical unit of the UK. IMD scores were attributed to all individuals in the sample based on their postcode and aggregated to quintiles (1 = most deprived, 5 = least deprived).

*Other measures*

Marital status (married and living with partner, not married and living with partner, not living with partner) was a control variable.

Data analysis

Respondents were matched to the park that they listed as their most visited in the summer months. Almost all participants (97.74%) responded to this question , and so this allowed the greatest number of respondents to be retained for analysis. Two parks were dropped during this process, as they were not reported by any participant as their most visited park in the summer months.

First, linear regression analysis was performed to identify the significant predictors of park satisfaction and park use from the park feature domains, size and proximity. Next, separate multilevel models were performed to investigate the relationship between park features and (1) park satisfaction and (2) park use. Two levels were included: individuals at the first level and parks at the second level. We included a park identifier as a random intercept; models were also tested with random slopes for each of the park features but this did not improve the model fit and so were not included. Model 1 included park-level variables identified as significant in the linear regression analyses only. Control variables were then entered sequentially to adjust for proximity (model 2), ethnicity (model 3), socioeconomic status and demographics (model 4: maternal education, financial status, marital and cohabitation status), and then IMD (model 5). Coefficients are interpreted in the same way as a single level regression model - the effect of a 1 unit increase in the explanatory variable on, in this instance, level of park satisfaction or minutes of park use.

We tested ethnicity and socioeconomic status as moderators of the relationship between park features and (1) park satisfaction and (2) park use by entering interaction terms in an unadjusted model. The results of the prior linear regression analyses determined which park features were tested alongside ethnicity, education, financial status and IMD quintile. Both main effects were entered as well as the interaction. We statistically tested interactions using the likelihood ratio test, compared to a model with no interaction term.

Mediation of the relationship between park features and park use by park satisfaction was evaluated using the Baron and Kenny approach (Baron and Kenny, 1986). Using this method, the following relationships are assessed: (1) the relationship between park features and park use, (2) the relationship between park features and park satisfaction, and (3) the association between park satisfaction and park use, and finally (4) the association between park features, park satisfaction and park use. Following this approach, only park feature domains that were identified as significant predictors of both park satisfaction and park use in the initial linear regression analyses were considered as the independent variable. To account for the clustered nature of the data, we used the multilevel mediation 'ml_mediation' package in Stata (version 14) (StataCorp, 2015). Bootstrapping was used to create standard errors (SEs) and 95% confidence intervals (95% CIs). Statistical significance was set at p-value ≤0.05. All analyses were carried out in Stata 14.

# 3 Results

Participants

The socio-demographics of participants are reported in Table 2. Almost half (47%) of the sample was Pakistani, with 38% White British and other ethnicities making up 15%. The sample was evenly split in terms of educational background. The majority reported they were not struggling financially (70%) and 71% reported they were married and living with a partner. Most of the sample were in the most or second-most deprived IMD quintile.

The sample had a mean (M) park satisfaction score of 2.15 with a standard deviation (SD) of 0.85. A one-way analysis of variance (ANOVA) was calculated on park satisfaction for all individual variables. Significant differences were observed between ethnicities, $F(2, 649) = 3.92$, $p = 0.02$; and between IMD quintiles, $F(4, 647) = 2.89$, $p = 0.02$. No other differences were observed. Post-hoc Tukey tests were conducted on all possible pairwise contrasts. In terms of differences between ethnicities, White British respondents (M = 2.26, SD = 0.83) reported significantly higher ($p = 0.01$) park satisfaction than Pakistani respondents (M = 2.06, SD = 0.85). Those in the least deprived IMD quintile (M = 2.06, SD = 0.88) reported significantly higher ($p = 0.03$) park satisfaction than those in the most deprived IMD quintile (M = 2.56, SD = 0.70). All other comparisons were not significant.

ANOVAs were also carried out to explore differences in park use. Significant differences were observed by ethnicity $F(2, 649) = 5.29$, $p = 0.005$ and marital status $F(2,649) = 6.85$, $p = 0.001$. Post-hoc Tukey tests revealed that the White British group spent (M = 272.89, SD = 267.45) significantly more time ($p = 0.004$) at the park than the Pakistani group (M = 207.45, SD = 212.31). Furthermore, persons not living with a partner (M = 279.05, SD = 248.47) spent significantly more time ($p = 0.004$) at the park than those who are married and living with partner (M = 214.15, SD = 225.12).

Linear regression analysis

Linear regression analyses were carried out to identify park features that predicted park satisfaction and park use (see Tables 3 and 4). Table 3 shows that the amenities, incivilities and usability domains significantly predicted park satisfaction. A higher amenities and usability domain score was associated with a higher park satisfaction score, whereas the presence of more incivilities was associated with reduced park satisfaction. Table 4 indicates incivilities were also negatively associated with park use, with weekly duration of use reduced by 23 minutes on average. Increased access was also associated with reduced park use. More natural features and greater size of the park were positively related to park use.

Multilevel modelling

*Park satisfaction*

A null model was fitted initially to assess whether the parks differ from each other, on average, on satisfaction scores (data not reported). A substantial proportion of the total variance in the park satisfaction score is accounted for by differences between parks (ICC = 23.55%). Adjusted multilevel models of park satisfaction are reported in Table 5. Variation drops considerably in model 1 when adding the park feature domains (model 1 ICC= 2.20%) and remains low in the fully adjusted model (model 5 ICC = 2.07%).

The table shows small but significant associations between park features and park satisfaction across all models. In the fully adjusted model, amenities and usability were positively related to park satisfaction (B = 0.07, 95% CI 0.01 to 0.13; B = 0.09, 95% CI 0.01 to 0.16), indicating that a 1-point increase in the amenity and usability domain scores was associated with a 0.07 and 0.09 increase in park satisfaction ratings respectively. Incivilities showed a negative association (B = -0.11, 95% CI 0.16 to 0.06). No significant associations were identified between park satisfaction and individual characteristics.

*Park use*

A null model was fitted to assess whether the parks differ from each other, on average, on duration of park use (data not reported). A small proportion of the total variance in parks use is accounted for by differences between parks (ICC = 7.57%). Adjusted multilevel models of park use are reported in Table 6. Variation drops when the park-level variables are added (ICC = 1.77%). This is further reduced in the fully adjusted model (ICC = 0.06%).

The fully adjusted model shows access (B= -115.19, 95% CI -183.54 to -46.83) and incivilities (B= -21.28, 95% CI -35.41 to 7.16) are significantly negatively associated with park use. This indicates that a 1-point increase in the access and incivilities domain scores was associated with a reduction in average weekly park use of 115 minutes and 21 minutes respectively. These patterns are consistent across models. There is also a marginal negative relationship between proximity and use (B= -0.01, 95% CI -0.02 to 0.002). Further, those not married and living with a partner, and those not living with a partner were associated with increased park use (B = 69.46, 95% CI 12.73 to 126.18, B= 70.19 95% CI 12.01 to 128.37).

*Does ethnicity or socioeconomic status have a moderating role?*

We then explored whether ethnicity or socioeconomic status moderated the relationship between park features, park satisfaction and park use. Following the results of the linear regressions, we tested interactions between amenities, incivilities and usability, and ethnicity and socioeconomic status, for park satisfaction; for the park use model we tested interactions between access, natural features, incivilities and size and ethnicity and socioeconomic status. We entered both main effects and tested each interaction separately, using a likelihood ratio test for significance. No statistically significant interactions were observed been park features, park satisfaction and park use.

*Does park satisfaction mediate the relationship between park features and park use?*

We then explored whether park satisfaction might mediate the relationship between park features and park use using multilevel mediation. Since the incivilities domain alone was significantly associated both with park satisfaction and park use, mediation was tested with this domain as the independent variable only. It was found that park satisfaction was not significantly associated with park use when controlling for incivilities. Further, bootstrapping confirmed that the indirect effect of incivilities on park use via park satisfaction was not significant (B =-3.28, 95% CI -7.00 to 0.43).

**4 Discussion**

This study aimed to examine the associations between park features, satisfaction and use. Amenities, incivilities and usability were found to be related to park satisfaction in the expected directions; size and proximity were not related. Access, incivilities and proximity were found to be significantly related to park use, although the effect of proximity was negligible. Further analyses revealed ethnicity and socioeconomic status were not moderators of the relationship between park features, satisfaction and use, and there was no evidence of mediation between park features and use by satisfaction.

The importance of amenities and the variety of activities available for encouraging park use has been demonstrated in previous research. For example, Edwards et al. (2015) audited 58 parks and surveyed 1304 adolescents in Western Australia, and identified features such as picnic tables and toilets, among others, to be associated with park use. Baran et al. (2014) also found shelters and picnic areas were positively related with park use in the US. However, this study attributed these features to park satisfaction, rather than park use. Furthermore, many more park features have been shown to be associated with park use that were not found in this study, such as playgrounds, table tennis tables, basketball courts, ponds and trees (Baran et al., 2014; Edwards et al., 2015; Park, 2019; Veitch et al., 2016). In this way, the evidence on park features, satisfaction and use remains rather mixed.

In this study it was found that incivilities were associated with both park satisfaction and use. This is in line with previous research that has consistently shown that signs of disorder such as graffiti, litter and vandalism are discouraging for park use and park-based physical activity (Douglas et al., 2018; Knapp, Gustat, Darensbourg, Myers, & Johnson, 2019; McCormack et al., 2010). Moreover, parks with more incivilities are more likely to be seen as less safe, which in turns reduces the chances of park use (Costigan et al., 2017; Derose, Han, Williamson, & Cohen, 2018; Lapham et al., 2016). Lastly, access was negatively related to park use. In this study, access was defined by the number of entrance points and the presence of paths. It was noted that the parks with few entrance points and no paths were small, local parks that were enclosed, oriented around playground equipment, and designed for small children. Given the nature of the sample, it is suggested that this explains the negative relationship.

There was no evidence of moderation by ethnicity or socioeconomic status on the relationship between park features and park satisfaction or use. This goes against current research into variation in park use, which has typically shown differences by population subgroups (Ho et al., 2005; Kaczynski et al., 2014; Payne, Mowen, & Orsega-Smith, 2002). Further work may be worthwhile to explore the differences in preference for park features. There was also no evidence that park satisfaction was a mediator of the relationship between park features and park use. Limited research has shown that residents are more likely to use their nearby green space when they are more satisfied with neighbourhood appearance, the variety of green space and the amount of open space, and vice versa (Hadavi & Kaplan, 2016). Further research is required to explore this relationship.

Amongst the main strengths of the study was the sample of ethnically diverse women from a predominantly deprived area in the UK. This study therefore reports findings from an understudied group in a novel context. We were also able to assess a considerable number of parks using a quality assessment tool (NEST, Gidlow et al., 2018) that was found to be reliable between observers. However, several limitations are acknowledged. The study is cross-sectional in design, precluding causal inferences. The study also had a fairly small sample size (n=652). The composition of the sample means the findings may be generalised to similarly deprived and multi-ethnic areas, but the extent to which the findings can be generalised to a more affluent or less ethnically diverse area is limited. In addition, the survey from which the park use and park satisfaction variables were derived asked the participant which park their *child* used and how satisfied they were with it. It may be that respondents visit other parks more frequently without their child, however, given the young age of their child it is suggested this is not likely.

Based on these findings, a number of recommendations can be offered to policymakers, urban health professionals and park managers who are looking to encourage park use. It is suggested that as the incivilities domain was associated with both park satisfaction and park use, the reduction of existing incivilities should be prioritised for intervention over the installation of new features. This may be strengthened by developing a 'monitoring' presence in the park, such as increasing park ranger presence or establishing or building park-based community groups. Amenities and usability were also related to park satisfaction, and therefore the maintenance or addition of items within these domains should be referred to when increasing satisfaction is the objective. Parks that are largely enclosed might also be promoted where the aim is to address safety concerns for parents of young children.

## 5 Conclusion

This study contributes to the limited research examining the associations between park features, satisfaction and use in an ethnically diverse and deprived sample in the UK. Varied park features were identified as being associated with park satisfaction and park use, including access, amenities, incivilities, usability and proximity to the park. Incivilities were significantly related to both satisfaction and use, suggesting that this feature is prioritised when designing future interventions.

**Acknowledgements**

We wish to thank Mark Ferguson, who assisted with data collection.

**Funding**

This research did not receive any specific grant from funding agencies in the public, commercial, or not-for-profit sectors.

**Declaration of interest**

None.

**Conflict of interest**

The authors have no conflict of interest to declare.

**References**

Baran, P. K., Smith, W. R., Moore, R. C., Floyd, M. F., Bocarro, J. N., Cosco, N. G., & Danninger, T. M. (2014). Park Use Among Youth and Adults: Examination of Individual, Social, and Urban Form Factors. *Environment and Behavior*, *46*(6), 768–800. https://doi.org/10.1177/0013916512470134

Barbosa, O., Tratalos, J. A., Armsworth, P. R., Davies, R. G., Fuller, R. A., Johnson, P., & Gaston, K. J. (2007). Who benefits from access to green space? A case study from Sheffield, UK. *Landscape and Urban Planning*, *83*(2–3), 187–195. https://doi.org/10.1016/j.landurbplan.2007.04.004

Bedimo-Rung, A. L., Mowen, A. J., & Cohen, D. A. (2005). The Significance of Parks to Physical Activity and Public Health A Conceptual Model. *American Journal of Preventive Medicine*, *28*(2S2), 159–168. https://doi.org/10.1016/j.ampre.2004.10.024

Bowler, D. E., Buyung-Ali, L. M., Knight, T. M., & Pullin, A. S. (2010). A systematic review of evidence for the added benefits to health of exposure to natural environments. *BMC Public Health*, *10*(1), 456. https://doi.org/10.1186/1471-2458-10-456

Bruton, C. M., & Floyd, M. F. (2014). Disparities in Built and Natural Features of Urban Parks: Comparisons by Neighborhood Level Race/Ethnicity and Income. *Journal of Urban Health*, *91*(5). https://doi.org/10.1007/s11524-014-9893-4

CABE. (2010). *Urban green nation: Building the evidence base*. Retrieved from http://webarchive.nationalarchives.gov.uk/20110118110347/http://www.cabe.org.uk/files/urban-green-nation.pdf

Cohen, D. A., Han, B., Derose, K. P., Williamson, S., Marsh, T., Raaen, L., & McKenzie, T. L. (2016). The Paradox of Parks in Low-Income Areas: Park Use and Perceived Threats. *Environment and Behavior*, *48*(1), 230–245. https://doi.org/10.1177/0013916515614366

Cohen, D. A., Han, B., Nagel, C. J., Harnik, P., McKenzie, T. L., Evenson, K. R., … Katta, S. (2016). The First National Study of Neighborhood Parks. *American Journal of Preventive Medicine*, *51*(4), 419–426. https://doi.org/10.1016/j.amepre.2016.03.021

Coombes, E., Jones, A. P., & Hillsdon, M. (2010). The relationship of physical activity and overweight to objectively measured green space accessibility and use. *Social Science & Medicine*, *70*(6), 816–822. https://doi.org/10.1016/j.socscimed.2009.11.020

Costigan, S. A., Veitch, J., Crawford, D., Carver, A., & Timperio, A. (2017). A cross-sectional investigation of the importance of park features for promoting regular physical activity in parks.

*International Journal of Environmental Research and Public Health*, *14*(11). https://doi.org/10.3390/ijerph14111335

Department of Communities and Local Government. (2011). The English Indices of Deprivation 2010. Retrieved from https://assets.publishing.service.gov.uk/government/uploads/system/uploads/attachment_data/file/6320/1870718.pdf

Department for Communities and Local Government. (2015). The English Indices of Deprivation 2015. Retrieved from https://assets.publishing.service.gov.uk/government/uploads/system/uploads/attachment_data/file/465791/English_Indices_of_Deprivation_2015_-_Statistical_Release.pdf

Derose, K. P., Han, B., Williamson, S., & Cohen, D. A. (2018). Gender Disparities in Park Use and Physical Activity among Residents of High-Poverty Neighborhoods in Los Angeles. *Women's Health Issues*, *28*(1), 6–13. https://doi.org/10.1016/j.whi.2017.11.003

Douglas, J. A., Briones, M. D., Bauer, E. Z., Trujillo, M., Lopez, M., & Subica, A. M. (2018). Social and environmental determinants of physical activity in urban parks: Testing a neighborhood disorder model. *Preventive Medicine*, *109*(December 2017), 119–124. https://doi.org/10.1016/j.ypmed.2018.01.013

Edwards, N., Hooper, P., Knuiman, M., Foster, S., & Giles-Corti, B. (n.d.). *Associations between park features and adolescent park use for physical activity*. https://doi.org/10.1186/s12966-015-0178-4

Engelberg, J. K., Conway, T. L., Geremia, C., Cain, K. L., Saelens, B. E., Glanz, K., … Sallis, J. F. (2016). Socioeconomic and race/ethnic disparities in observed park quality. *BMC Public Health*, *16*(1), 395. https://doi.org/10.1186/s12889-016-3055-4

ESRI 2018. ArcGIS Desktop: Release 10.6.1. Redlands, CA: Environmental Systems Research Institute.

Evenson, K. R., Wen, F., Hillier, A., & Cohen, D. A. (2013). Assessing the contribution of parks to physical activity using global positioning system and accelerometry. *Medicine and Science in Sports and Exercise*, *45*(10), 1981–1987. https://doi.org/10.1249/MSS.0b013e318293330e

Ferguson, M., Roberts, H., McEachan, R. R. C., & Dallimer, M. (2018). Contrasting distributions of urban green infrastructure across social and ethnic groups. *Landscape and Urban Planning*.

Floyd, M. F., Taylor, W. C., & Whitt-Glover, M. (2009). Measurement of park and recreation environments that support physical activity in low-income communities of color: Highlights of challenges and recommendations. *American Journal of Preventive Medicine*, *36*(4, Suppl),

S156–S160. Retrieved from http://ovidsp.ovid.com/ovidweb.cgi?T=JS&CSC=Y&NEWS=N&PAGE=fulltext&D=psyc6&AN=2009-03790-003

Gascon, M., Triguero-Mas, M., Martínez, D., Dadvand, P., Forns, J., Plasència, A., & Nieuwenhuijsen, M. (2015). Mental Health Benefits of Long-Term Exposure to Residential Green and Blue Spaces: A Systematic Review. *International Journal of Environmental Research and Public Health*, *12*(4), 4354–4379. https://doi.org/10.3390/ijerph120404354

Gidlow, C., van Kempen, E., Smith, G., Triguero-Mas, M., Kruize, H., Gražulevičienė, R., … Nieuwenhuijsen, M. J. (2018). Development of the natural environment scoring tool (NEST). *Urban Forestry and Urban Greening*, *29*(April 2017), 322–333. https://doi.org/10.1016/j.ufug.2017.12.007

Gobster, P. H. (2002). Managing Urban Parks for a Racially and Ethnically Diverse Clientele. *Leisure Sciences*, *24*, 143–159. Retrieved from https://www.nrs.fs.fed.us/pubs/jrnl/2002/nc_2002_Gobster_002.pdf

Hadavi, S., & Kaplan, R. (2016). Neighborhood satisfaction and use patterns in urban public outdoor spaces: Multidimensionality and two-way relationships. *Urban Forestry and Urban Greening*, *19*, 110–122. https://doi.org/10.1016/j.ufug.2016.05.012

Hartig, T., Mitchell, R., de Vries, S., & Frumkin, H. (2014). Nature and Health. *Annual Review of Public Health*, *35*(1), 207–228. https://doi.org/10.1146/annurev-publhealth-032013-182443

Ho, C., Sasidharan, V., Elmendorf, W., Willits, F. K., Graefe, A., & Godbey, G. (2005). Gender and Ethnic Variations in Urban Park Preferences, Visitation, and Perceived Benefits. *Journal of Leisure Research*, *37*(1), 281–306.

Hughey, S. M., Walsemann, K. M., Child, S., Powers, A., Reed, J. A., & Kaczynski, A. T. (2016). Using an environmental justice approach to examine the relationships between park availability and quality indicators, neighborhood disadvantage, and racial/ethnic composition. *Landscape and Urban Planning*, *148*, 159–169. https://doi.org/10.1016/j.landurbplan.2015.12.016

Jennings, V., Johnson Gaither, C., & Gragg, R. S. (2012). Promoting Environmental Justice Through Urban Green Space Access: A Synopsis. *Environmental Justice*, *5*(1), 1–7. https://doi.org/10.1089/env.2011.0007

Kaczynski, A. T., Besenyi, G. M., Wilhelm Stanis, S. A., Javad Koohsari, M., Oestman, K. B., Bergstrom, R., … Reis, R. S. (2014). Are park proximity and park features related to park use and park-based physical activity among adults? Variations by multiple socio-demographic characteristics. *International Journal of Behavioral Nutrition and Physical Activity*, *11*. https://doi.org/10.1186/s12966-014-0146-4

Kaczynski, A. T., Potwarka, L. R., & Saelens, B. E. (2008). Association of Park Size, Distance, and Features With Physical Activity in Neighborhood Parks. *American Journal of Public HealthAm J Public Health*, *9898*(810), 1451–1456. https://doi.org/10.2105/AJPH.2007.129064

Kaczynski, A. T., Potwarka, L. R., Smale, B. J. A., & Havitz, M. F. (2009). Association of Parkland proximity with neighborhood and park-based physical activity: Variations by gender and age. *Leisure Sciences*, *31*(2), 174–191. https://doi.org/10.1080/01490400802686045

Kessel, A., Green, J., Pinder, R., Wilkinson, P., Grundy, C., & Lachowycz, K. (2009). Multidisciplinary research in public health: A case study of research on access to green space. *Public Health*, *123*(1), 32–38. https://doi.org/10.1016/j.puhe.2008.08.005

Knapp, M., Gustat, J., Darensbourg, R., Myers, L., & Johnson, C. (2019). The relationships between park quality, park usage, and levels of physical activity in low-income, African American neighborhoods. *International Journal of Environmental Research and Public Health*, *16*(1). https://doi.org/10.3390/ijerph16010085

Lapham, S. C., Cohen, D. A., Han, B., Williamson, S., Evenson, K. R., McKenzie, T. L., … Ward, P. (2016). How important is perception of safety to park use? A four-city survey. *Urban Studies*, *53*(12), 2624–2636. https://doi.org/10.1177/0042098015592822

Maas, J., Verheig, R. A., Groenewegen, P. P., de Vries, S., & Spreeuwenberg, P. (2006). Green space, urbanity, and health: how strong is the relation? *Journal of Epidemiology & Community Health*, *60*(7), 587–592. https://doi.org/10.1136/jech.2005.043125

McCormack, G. R., Rock, M., Toohey, A. M., & Hignell, D. (2010). Characteristics of urban parks associated with park use and physical activity: A review of qualitative research. *Health and Place*, *16*(4), 712–726. https://doi.org/10.1016/j.healthplace.2010.03.003

Mitchell, R., & Popham, F. (2007). Greenspace, urbanity and health: Relationships in England. *Journal of Epidemiology and Community Health*, *61*(8), 681–683. Retrieved from http://ovidsp.ovid.com/ovidweb.cgi?T=JS&CSC=Y&NEWS=N&PAGE=fulltext&D=psyc5&AN=2007-11811-005

Mitchell, R., & Popham, F. (2008). Effect of exposure to natural environment on health inequalities: an observational population study. *The Lancet*, *372*(9650), 1655–1660. https://doi.org/10.1016/S0140-6736(08)61689-X

Nieuwenhuijsen, M. J., Khreis, H., Triguero-Mas, M., Gascon, M., & Dadvand, P. (2017). Fifty Shades of Green: Pathway to Healthy Urban Living. *Epidemiology*, *28*(1), 63–71. https://doi.org/10.1097/EDE.0000000000000549

Office for National Statistics ; National Records of Scotland ; Northern Ireland Statistics and Research Agency (2017): 2011 Census aggregate data. UK Data Service (Edition: February 2017). DOI: http://dx.doi.org/10.5257/census/aggregate-2011-2

Park, K. (2019). Park and Neighborhood Attributes Associated With Park Use: An Observational Study Using Unmanned Aerial Vehicles. *Environment and Behavior*, 001391651881141. https://doi.org/10.1177/0013916518811418

Payne, L. L., Mowen, A. J., & Orsega-Smith, E. (2002). An Examination of Park Preferences and Behaviors Among Urban Residents: The Role of Residential Location, Race, and Age. *Leisure Sciences*, *24*(2), 181–198. https://doi.org/10.1080/01490400252900149

Richardson, E. A., Mitchell, R., Hartig, T., Vries, S. De, Astell-Burt, T., & Frumkin, H. (2012). Green cities and health: a question of scale? *Journal of Epidemiology & Community Health*, *66*, 160–165. https://doi.org/10.1136/jech.2011.137240

Rigolon, A., Browning, M., & Jennings, V. (2018). Inequities in the quality of urban park systems: An environmental justice investigation of cities in the United States. *Landscape and Urban Planning*, *178*(May), 156–169. https://doi.org/10.1016/j.landurbplan.2018.05.026

Roe, J., Aspinall, P. A., & Thompson, C. W. (2016). Understanding relationships between health, ethnicity, place and the role of urban green space in deprived urban communities. *International Journal of Environmental Research and Public Health*, *13*(7), 1–21. https://doi.org/10.3390/ijerph13070681

Rung, A. L., Mowen, A. J., Broyles, S. T., & Gustat, J. (2011). The role of park conditions and features on park visitation and physical activity. *Journal of Physical Activity & Health*, *8 Suppl 2*(Suppl 2), S178-87. https://doi.org/10.1123/jpah.8.s2.s178

Shores, K. A., & West, S. T. (2008). The relationship between built park environments and physical activity in four park locations. *Journal of Public Health Management and Practice : JPHMP*, *14*(3), e9-16. https://doi.org/10.1097/01.PHH.0000316495.01153.b0

StataCorp. 2015. Stata Statistical Software: Release 14. College Station, TX: StataCorp LP.

Suminski, R. R., Connolly, E. K., May, L. E., Wasserman, J., Olvera, N., & Lee, R. E. (2012). Park quality in racial/ethnic minority neighborhoods. *Environmental Justice*, *5*(6), 271–278.

Timperio, A., Ball, K., Salmon, J., Roberts, R., & Crawford, D. (2007). Is availability of public open space equitable across areas? *Health and Place*, *13*(2), 335–340. https://doi.org/10.1016/j.healthplace.2006.02.003

Twohig-Bennett, C., & Jones, A. (2018). The health benefits of the great outdoors: A systematic review and meta-analysis of greenspace exposure and health outcomes. *Environmental Research*, *166*(June), 628–637. https://doi.org/10.1016/j.envres.2018.06.030

van Hecke, L., Verhoeven, H., Clarys, P., Dyck, D., van de Weghe, N., Baert, T., … van Cauwenberg, J. (2018). Factors related with public open space use among adolescents: A study using GPS and accelerometers. *International Journal of Health Geographics*, *17*(1), 1–16. https://doi.org/10.1186/s12942-018-0123-2

Vaughan, C. A., Colabianchi, N., Hunter, G. P., Beckman, R., & Dubowitz, T. (2018). Park Use in Low-Income Urban Neighborhoods: Who Uses the Parks and Why? *Journal of Urban Health*, *95*(2), 222–231. https://doi.org/10.1007/s11524-017-0221-7

Vaughan, K. B., Kaczynski, A. T., Stanis, S. A. W., Besenyi, G. M., Bergstrom, R., & Heinrich, K. M. (2013). *Exploring the Distribution of Park Availability, Features, and Quality Across Kansas City, Missouri by Income and Race/Ethnicity: an Environmental Justice Investigation*. https://doi.org/10.1007/s12160-012-9425-y

Veitch, J., Salmon, J., Parker, K., Bangay, S., Deforche, B., & Timperio, A. (2016). Adolescents' ratings of features of parks that encourage park visitation and physical activity. *International Journal of Behavioral Nutrition and Physical Activity*, *13*(1), 1–10. https://doi.org/10.1186/s12966-016-0391-9

WHO Regional Office for Europe. (2016). Urban green spaces and health: a review of the evidence. *World Health Organization*, 1:92. Retrieved from http://www.euro.who.int/__data/assets/pdf_file/0005/321971/Urban-green-spaces-and-health-review-evidence.pdf?ua=1

Wolch, J. R., Byrne, J., & Newell, J. P. (2014). Urban green space, public health, and environmental justice: The challenge of making cities "just green enough." *Landscape and Urban Planning*, *125*, 234–244. https://doi.org/10.1016/j.landurbplan.2014.01.017

Wright, J., Small, N., Raynor, P., Tuffnell, D., Bhopal, R., Cameron, N., … West, J. (2013). Cohort Profile: the Born in Bradford multi-ethnic family cohort study. *International Journal of Epidemiology*, *42*(4), 978–991. https://doi.org/10.1093/ije/dys112

**List of Figures**

**Figure 1.** The location of the audited parks, shown in green (left). The right hand panel shows the location of Bradford in the UK.

**Figure 2.** Flow diagram of participant selection

**Figure 1.** The location of the audited parks, shown in green (left). The right-hand panel shows the location of Bradford in the UK.

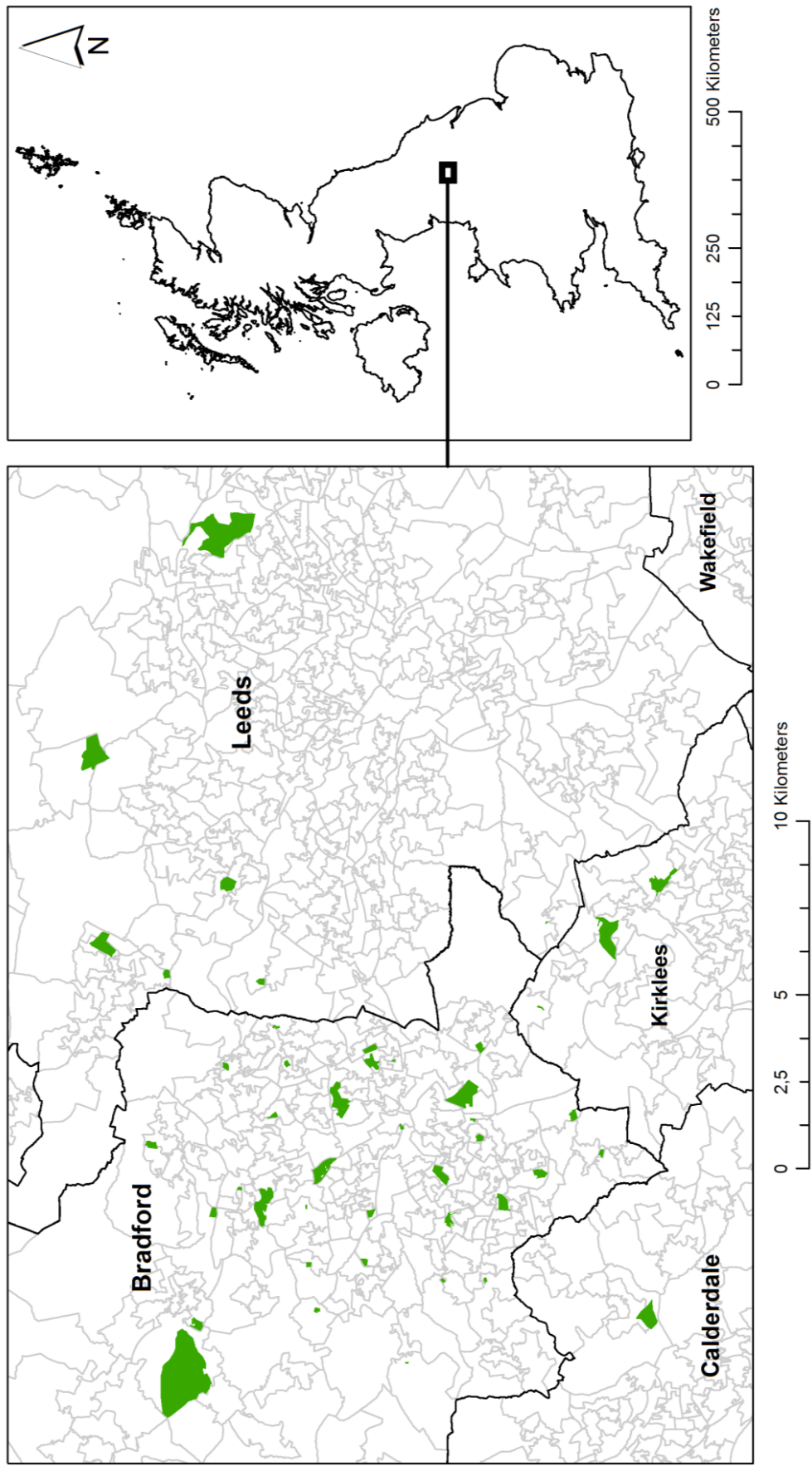

**Figure 2.** Flow diagram of participant selection

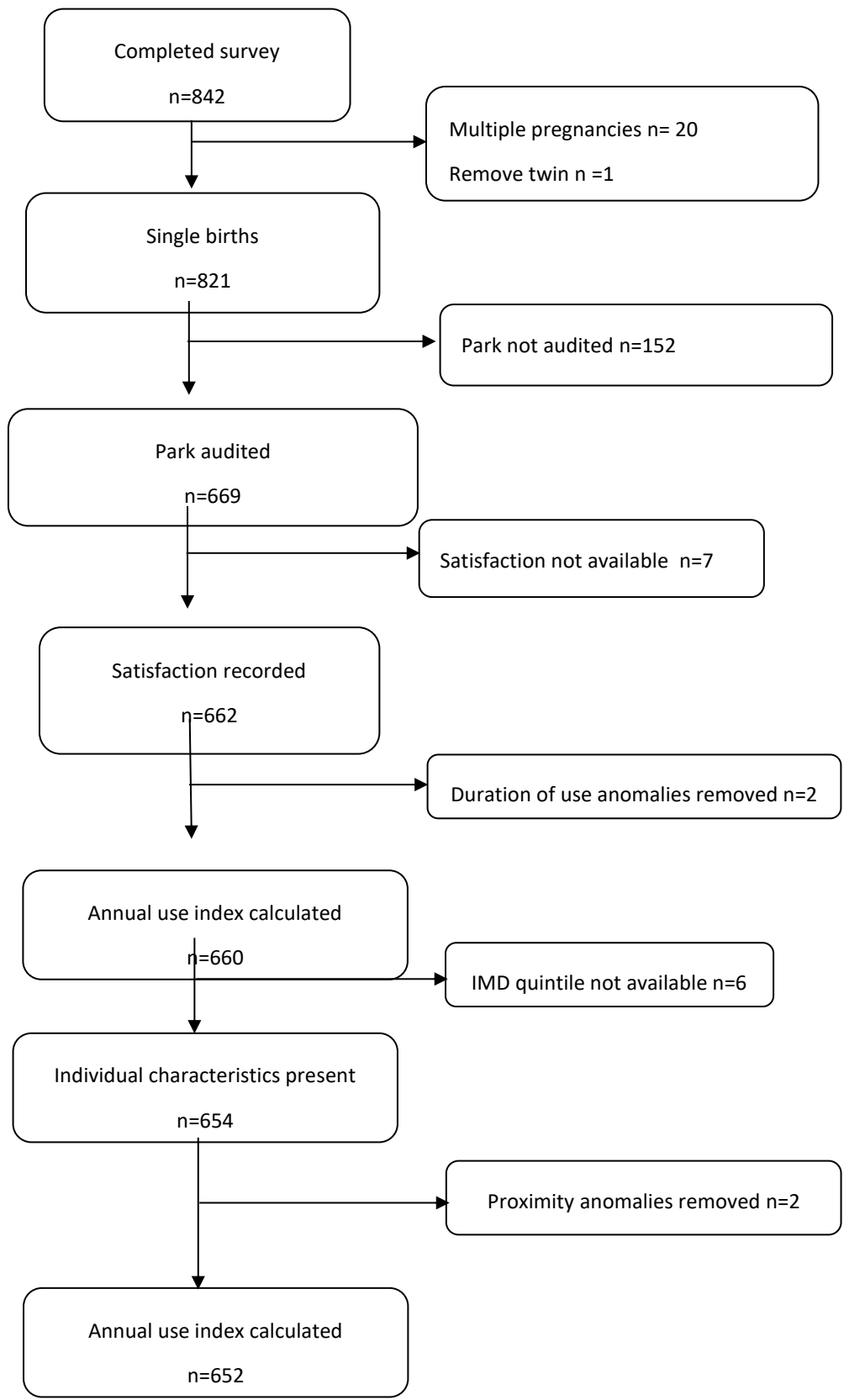

**Table 1.** Park features by domain

| Domain | Features recorded for presence |
| --- | --- |
| Access | Entrance points, walking/ cycling paths |
| Recreational Facilities | Playground equipment, grass pitches, courts (e.g. tennis, basketball), skateboard ramps, other sports or fitness facilities, presence of open space |
| Amenities | Seating/benches, litter bins, dog mess bins, public toilets, café/kiosk, man-made shelter, picnic tables, drinking fountains |
| Aesthetics – Natural features | Flower beds, planters or wild flowers; other planted trees, shrubs or plants |
| Aesthetics – Non-natural features | Water fountain, other public art, historic or attractive buildings or other man-made structures |
| Incivilities | General litter, evidence of alcohol use, evidence of drug taking, graffiti, broken glass, vandalism, dog mess, excessive/ unpleasant noise, unpleasant smells |
| Significant Natural Features | Water features, good view points, vistas, scenic views; trees |
| Usability | Sport, informal games, walking/running, children's play, conservation/biodiversity, enjoying the landscape/ visual qualities, meeting or socialising with friends/ neighbours, relaxing/ unwinding, cycling |

**Table 2.** Characteristics of study participants

|  | Total sample n (%) | Satisfaction score= 1 (lowest) n (%) | Satisfaction score= 2 n (%) | Satisfaction score= 3 (highest) n (%) | Park use (mins) M(SD) |
|---|---|---|---|---|---|
| Full sample | 652 (100) | 190 (29.14) | 169 (25.92) | 293 (44.94) | 235.57 (237.22) |
| Ethnicity | | | | | |
| White British | 245 (37.58) | 61 (24.90) | 58 (23.67) | 126 (51.43) | 272.89 (267.45) |
| Pakistani | 309 (47.39) | 102 (33.01) | 85 (27.51) | 122 (39.48) | 207.45 (212.31) |
| Other | 98 (15.03) | 27 (27.55) | 26 (26.53) | 45 (45.92) | 230.91 (220.36) |
| Education | | | | | |
| Maximum of 5 GCSEs | 321 (49.23) | 99 (30.84) | 72 (22.43) | 150 (46.73) | 225.42 (222.77) |
| A level equivalent | 331 (50.77) | 91 (27.49) | 97 (29.31) | 143 (43.20) | 245.40 (250.39) |
| Financial status | | | | | |
| Struggling financially | 199 (30.52) | 60 (30.15) | 55 (27.64) | 84 (42.21) | 236.62 (221.90) |
| Not struggling financially | 453 (69.48) | 130 (28.70) | 114 (25.17) | 209 (46.14) | 235.10 (243.88) |
| Marital status | | | | | |
| Married and living with partner | 465 (71.32) | 137 (29.46) | 124 (26.67) | 204 (43.87) | 214.15 (225.12) |
| Not married and living with partner | 105 (16.10) | 24 (22.86) | 26 (24.76) | 55 (52.38) | 296.44 (266.04) |
| Not living with partner | 82 (12.58) | 29 (35.37) | 19 (23.17) | 34 (41.46) | 279.05 (248.47) |
| IMD quintile | | | | | |
| 1 (most deprived) | 233 (35.74) | 82 (35.19) | 53 (22.75) | 98 (42.06) | 205.29 (198.65) |
| 2 | 171 (26.23) | 50 (29.24) | 44 (25.73) | 77 (45.03) | 232.28 (216.83) |
| 3 | 126 (19.33) | 26 (20.63) | 39 (30.95) | 61 (48.41) | 280.43 (298.83) |
| 4 | 95 (14.57) | 29 (30.53) | 27 (28.42) | 39 (41.05) | 250.58 (250.21) |
| 5 (least deprived) | 27 (4.14) | 3 (11.11) | 6 (22.22) | 18 (66.67) | 255.46 (273.00) |

**Table 3**. Linear regression of park characteristics on park satisfaction (n=42)

|  | B (95% CI) |
|---|---|
| Access | -0.01 (-0.22, 0.23) |
| Recreational facilities | -0.03 (-0.11 0.04) |
| Amenities | 0.08 (0.02, 0.14)* |
| Natural features | -0.08 (-0.23, 0.06) |
| Non-natural features | 0.002 (-0.10, 0.10) |
| Significant natural features | -0.04 (-0.22, 0.14) |
| Incivilities | -0.13 (-0.17, -0.08)*** |
| Usability | 0.14 (0.04, 0.23)** |
| Size | -0.001 (-0.003, 0.01) |

*$p < 0.05$ ** $p < 0.01$ *** $p < 0.001$

$F_{(9, 642)} = 18.48$ with an $R^2$ of 0.2058 (adjusted $R^2$ 0.1947)

Table 4. Linear regression of park characteristics on average weekly park use (mins) (n=652)

| | B (95% CI) |
|---|---|
| Access | -87.36 (-157.24, -17.48)* |
| Recreational facilities | 1.90 (-20.16, 23.95) |
| Amenities | -4.66 (-23.07, 13.74) |
| Natural features | 45.96 (0.92, 91.01)* |
| Non-natural features | -6.00 (-36.30, 24.29) |
| Significant natural features | -37.78 (-92.01, 16.45) |
| Incivilities | -22.96 (-37.25, -8.67)* |
| Usability | -19.35 (-48.15, 9.44) |
| Size | 0.59 (0.05, 1.15)* |

*p < 0.05 ** p < 0.01 ***p < 0.001

$F(9, 642) = 5.67$ with an $R^2$ of 0.0736 (adjusted $R^2$ 0.0607)

**Table 5.** Multilevel models for effects of NEST domains (model 1) and socioeconomic and demographic information (model 2-5) on park satisfaction

|  | Model 1 Adjusted for park variables | Model 2 Adjusted for proximity | Model 3 Adjusted for ethnicity | Model 4 Adjusted for SES and demographics | Model 5 Adjusted for IMD quintile |
|---|---|---|---|---|---|
| Amenities | 0.07 (0.01, 0.13)* | 0.06 (0.003, 0.12)* | 0.06 (0.003, 0.13)* | 0.06 (0.002, 0.12)* | 0.07 (0.01, 0.13)* |
| Incivilities | -0.12 (-0.17, -0.08)*** | -0.12 (-0.17, -0.07)*** | -0.11 (-0.16, -0.07)*** | -0.11 (-0.16, -0.06)*** | -0.11 (-0.16, -0.06)*** |
| Usability | 0.09 (0.03, 0.20)* | 0.09 (0.01, 0.17)* | 0.09 (0.01, 0.17)* | 0.09 (0.01, 0.17)* | 0.09 (0.01, 0.16)* |
| Proximity |  | -0.00001 (-0.0001, 0.0004) | -0.00001 (-0.00001, 0.00004) | -0.00001 (-0.00001, 0.00004) | -0.00001 (-0.0000, 0.0000) |
| **Ethnicity** |  |  |  |  |  |
| Pakistani |  |  | -0.06 (-0.21, 0.09) | -0.07 (-0.24, 0.10) | -0.06 (-0.24, 0.12) |
| Other |  |  | -0.02 (-0.21, 0.16) | -0.02 (-0.21, 0.17) | -0.02 (-0.22, 0.18) |
| **Education** |  |  |  |  |  |
| A level equivalent or higher |  |  |  | -0.06 (-0.18, 0.06) | -0.07 (-0.19, 0.05) |
| **Financial status** |  |  |  |  |  |
| Not struggling financially | | | | -0.02 (-0.11, 0.14) | -0.02 (-0.11, 0.15) |
| **Marital status** | | | | | |
| Not married and living with partner | | | | 0.02 (-0.16, 0.21) | 0.03 (-0.16, 0.21) |
| Not living with partner | | | | -0.16 (-0.35, 0.03) | -0.15 (-0.34, 0.04) |
| **IMD quintile** | | | | | |
| 2 | | | | | 0.05 (-0.10, 0.20) |
| 3 | | | | | 0.10 (-0.07, 0.27) |
| 4 | | | | | -0.06 (-0.26, 0.13) |
| 5 | | | | | 0.16 (-0.17, 0.49) |
| Constant | 1.31 (0.65, 1.98) | 1.34 (0.67, 2.01) | 1.34 (0.68, 2.01) | 1.36 (0.68, 2.04) | 1.34 (0.65, 2.03) |
| **ICC (%)** | 2.20 | 2.01 | 2.16 | 2.26 | 2.07 |

\* p < 0.05, \*\*p < 0.01, \*\*\*p < 0.001

Unstandardised coefficient (B) and 95% CIs reported

**Table 6.** Multilevel models for effects of NEST domains (model 1) and socioeconomic and demographic information (model 2-5) on park use

|  | Model 1 Adjusted for park variables | Model 2 Adjusted for proximity | Model 3 Adjusted for ethnicity | Model 4 Adjusted for SES and demographics | Model 5 Adjusted for IMD quintile |
|---|---|---|---|---|---|
| Access | -114.50 (-186.16, -42.84)** | -116.20 (-187.26, -45.15)** | -120.49 (-188.49, -52.48)** | -119.84 (-187.37, -52.31)** | -115.19 (-183.54, -46.83)** |
| Natural | 8.18 (-32.19, 48.56) | 7.81 (-32.08, 47.70) | 5.50 (-31.34, 42.33) | 1.61 (-35.15, 38.37) | -0.52 (-37.50, 36.46) |
| Incivilities | -20.69 (-34.73, -6.65)** | -25.76 (-40.35, -11.17)** | -20.82 (-34.57, -7.07)** | -21.14 (-34.91, -7.37)** | -21.28 (-35.41, -7.16)* |
| Size | 0.18 (-0.41, 0.76) | 0.37 (-0.23, 0.98) | 0.39 (-0.16, 0.94) | 0.46 (-0.09, 1.02) | 0.50 (-0.06, 1.06) |
| Proximity |  | -0.01 (-0.17, 0.001)* | -0.01 (-0.02, 0.002)* | -0.01 (-0.02, -0.003)** | -0.01 (-0.02, -0.002)** |
| **Ethnicity** |  |  |  |  |  |
| Pakistani |  |  | -48.10 (-91.50, -4.70)* | -16.81 (-66.14, 32.51) | -13.94 (-66.70, 38.82) |
| Other |  |  | -29.40 (-85.10, 26.31) | -13.43 (-71.17, 44.30) | -14.41 (-74.52, 45.71) |
| **Education** |  |  |  |  |  |
| A level equivalent or higher |  |  |  | 31.94 (-4.68, 68.57) | 29.81 (-7.26, 66.90) |
| **Financial status** |  |  |  |  |  |
| Not struggling financially |  |  |  | -3.53 (-43.29, 36.23) | -2.16 (-42.01, 37.69) |
| **Marital status** | | | | | |
| Not married and living with partner | | | | 69.43 (12.68, 126.19) | 69.46 (12.73, 126.18) |
| Not living with partner | | | | 67.06 (9.46, 124.66) | 70.19 (12.01, 128.37) |
| **IMD quintile** | | | | | |
| 2 | | | | | 15.88 (-29.77, 61.53) |
| 3 | | | | | 46.01 (-6.80, 98.82) |
| 4 | | | | | 3.37 (-56.33, 63.08) |
| 5 | | | | | -22.993 (-123.27, 77.30) |
| Constant | 507.89 (362.28, 653.50) | 542.80 (395.18, 690.43) | 561.70 (418.76, 704.64) | 517.50 (370.07, 664.93) | 498.19 (340.33, 656.05) |
| **ICC (%)** | 1.77 | 1.63 | 0.05 | 0.05 | 0.06 |

\* p < 0.05, \*\*p < 0.01

Unstandardised coefficients (B) and 95% CIs reported